\documentclass[english,preprint]{aastex}
\usepackage[T1]{fontenc}
\usepackage[latin9]{inputenc}
\usepackage{geometry}
\usepackage{textcomp}
\usepackage{amssymb}
\usepackage{graphicx}
\usepackage{babel}
\begin{document}

\title{Noise-gating to clean astrophysical image data}

\author{C.E. DeForest}

\affil{Southwest Research Institute, 1050 Walnut Street, Boulder, CO, USA}
\begin{abstract}
I present a family of algorithms to reduce noise in astrophysical
images and image sequences, preserving more information from the original
data than is retained by conventional techniques. The family uses
locally adaptive filters (``noise gates'') in the Fourier domain,
to separate coherent image structure from background noise based on
the statistics of local neighborhoods in the image. Processing of
solar data limited by simple shot noise or by additive noise reveals
image structure not easily visible in the originals, preserves photometry
of observable features, and reduces shot noise by a factor of 10 or
more with little to no apparent loss of resolution, revealing faint
features that were either not directly discernible or not sufficiently
strongly detected for quantitative analysis. The method works best
on image sequences containing related subjects, for example movies
of solar evolution, but is also applicable to single images provided
that there are enough pixels. The adaptive filter uses the statistical
properties of noise and of local neighborhoods in the data, to discriminate
between coherent features and incoherent noise without reference to
the specific shape or evolution of the those features. The technique
can potentially be modified in a straightforward way to exploit additional a priori
knowledge about the functional form of the noise. 
\end{abstract}

\section{\label{sec:Introduction}Introduction}

Images and image sequences are central to astrophysics and especially to the
subfield of solar physics.
Most such images are affected by shot noise or other additive noise,
which limit sensitivity and, indirectly, spatial and temporal resolution
of the data: small or short-duration image features may not rise above
the noise floor even though they are, at least in principle, resolved
by the instrument that collected the image. 

The most commonly-used denoising methods involve smoothing the data
by direct convolution with a rectangular or smooth kernel, by median
filtering, or by fitting a semi-empirical model to relevant image
features. All of these methods remove more information from the noise
than is strictly necessary. For example, smoothing averages across
pixels to beat down the noise, but also discards all of the high resolution
(small-feature) aspects of the data. Semi-empirical fitting extracts
a few parameters from a larger data set, but at the cost of ignoring
all aspects of the data that do not match the structure of the fitted
model. 

In general, noise reduction and detection schemes attempt to isolate
aspects of the data that contain mainly noise, from those aspects
of the data that contain signal. Convolutional smoothing is an example:
it works because the Fourier transform concentrates image structure
near the origin, while keeping additive noise spread across the entire
Fourier domain. Convolution with a smoothing kernel attenuates the
high frequencies, reducing the overall Fourier energy of the noise
spectrum \textendash{} at the cost of discarding any information contained
in those high frequencies (e.g., \citealt{bracewell_fourier_2000}). 

But noise, despite being a random variable, has clearly defined statistical
characteristics, that can be used to separate it from signal on the
basis of coherence within a small neighorhood of pixels (e.g., \citealt{lee_digital_1980}).
Adaptive filters that take advantage of these statistics by with characteristics
dependent on local image characteristics are well studied in the image
processing literature (two good reviews are \citealt{shynk_frequency-domain_1992}
and \citealt{buades_review_2005}). Adaptive filters have long been
used to take advantage of known functional dependence between image
values and noise level (\citealt{kuan_adaptive_1985}); such filters
remain a topic of active research and refinement (\citealt{shaick_hybrid_2000,huang_bayesian_2015,lebrun_noise_2015}).
A particularly useful class of adaptive image filter is a 2-D noise-gate:
a filter that identifies which Fourier components rise above a modeled
noise spectrum, and attenuates or discards those that are weaker while
retaining those that are strong. 

Noise-gating in the Fourier domain is a technique takes further advantage
of the properties of the Fourier transform. In particular, image patterns
that might be recognized as ``features'' typically have a concentrated
Fourier spectrum. Hence, in each small region of a scientific image
containing noisy but discernible structure, a few Fourier components
typically rise above the noise floor. This concentration can be used
to retain the ``features'' while attenuating or rejecting components
containing more noise. This process is more selective than direct
convolution by a single smoothing kernel, because the filter is locally
adapted to preserve identified signal in each neighborhood of the
image.

Noise gating is commonly used in 1-D, to ``clean'' audio signals
(\citealt{davis_sound_1989}). A time-series of pressure levels (audio)
is broken into segments, each of which is Fourier transformed. Fourier
components whose amplitudes rise above a predetermined noise floor
are retained, and all others are either zeroed or attenuated. Then
the segments are recombined to create the cleaned audio signal. The
technique works to remove hum, buzz, crowd noise, tape hiss, and other
additive noise from high quality recordings.

Generalizing noise-gating to 2-D (image) or 3-D (image sequences)
allows far better discrimination between information and noise than
simple smoothing, simply by taking advantage of the redundancy inherent
in visually distinguishable image structure: as in single images,
coherent ``features'' tend to produce localized spatiotemporal spectra.
The practical result is that statistically significant features in
solar, astrophysical, or other image sequences can be retained even
at the full instrument resolution, while the background noise is strongly
attenuated by a zero-amplitude gate, a simple coefficient, or a component-wise
adjustable Wiener filter.

Noise-gating in 2-D has been used since the mid 1990s to process still
images with additive noise (e.g., \citealt{yaroslavsky_local_1996}),
but 3-D applications have largely relied on separate spatial and temporal
filtering, using local motion tracking in video to allow co-adding
with minimal motion blur. These deblurring techniques do not work
well for images (such as solar magnetograms) that do not contain conventional
objects as would a cinematographic scene, or where the direction of
optical flow is ambiguous (\citealt{guo_temporal_2007,jovanov_combined_2009,liu_high-quality_2010}).
Generalizing the Fourier noise gate to 3-D affords the same basic
benefit as the transition to 2-D noise-gating: it helps to discriminate
coherent structure in both space and time, without regard for the
particular shape or evolution of the coherent pattern itself, thereby
preventing the need to track motion and allowing denoising of astrophysical
imagery. 

Finally, local-neighborhood noise-gating permits adjusting the noise
model based on local image characteristics. This permits gating noise
spectra that vary with location or with image characteristics, for
example shot noise from a telescope with a non-flat vignetting function.
Image and image-sequence noise gating are sufficiently promising and
spectacular, compared to smoothing, that they should be standard parts
of the data reduction ``toolbox'' used for heliophysical and astrophysical
image data. In this article, I present a method for noise-gating images
and image sequences with variable (shot) or constant (fixed additive)
noise spectra, and demonstrate its performance on commonly used image
sources in solar physics.

In Section \ref{sec:Background-=000026-Technique} I describe the
technique of noise-gating in detail. In Section \ref{sec:Results}
I present results from four data sets: shot-noise-limited image sequences
from \emph{SDO}/AIA; additive-noise-limited sequences from \emph{SDO}/HMI;
shot- and film-grain-limited still images from the NIXT sounding rocket;
and a standard template image used by the image processing community.
In Section \ref{sec:Discussion-=000026-Conclusions} I discuss the
results and their relevance to other data sets of astrophysical interest.
Finally, in Section \ref{sec:Conclusions} I summarize some advantages
and limitations of noise-gating and similar methods, for improving
the use of astrophysical and solar remote-sensing data.

\section{\label{sec:Background-=000026-Technique}Noise-gating technique}

Monochrome image sequences are mappings $Im:\mathbb{Z}^{3}\rightarrow\mathbb{R}$,
where the domain runs over pixel coordinates and the range describes
image brightness. In general, practical image data (after correction
for fixed detector effects, i.e. ``Flat-fielding'') contain at least
additive and shot noise:
\begin{equation}
Im(x,y,t)=Im_{0}(x,y,t)+N_{a}(x,y,t)+N_{s}(x,y,t)+N_{other}(x,y,t),\label{eq:image plus noise}
\end{equation}
 where $Im$ is an image sequence dataset, $Im_{0}$ is the ``ideal''
noise-free dataset, $N_{a}$ is an additive ``background noise''
term independent of $Im_{0}$, $N_{s}$ is an additive ``shot noise''
term that depends on $Im_{0}$, and $N_{other}$ is all other noise
sources. 

In high-fidelity audio processing and in some image applications,
$N_{a}$ is the dominant noise/background term. For example, in audio
data (which can be described as a single-pixel ``image'' sequence),
$N_{a}$ is commonly composed of white- or pink-noise ``static'',
highly spectrally peaked ``hum'', crowd noises, or other uniform
background noises that do not change character or amplitude across
time. This noise is commonly removed from high fidelity audio by ``noise-gating''.
The audio recording is broken into small segments that are individually
Fourier-transformed, and the Fourier amplitude spectrum of $N_{a}$
is determined from a segment that contains little or none of the actual
signal to be reconstructed (``active silence''). The silence can
be identified manually, or identified automatically by searching for
the minimum spectrum from a long recording that contains at least
one silent interval. Then all segments are processed by zeroing or
greatly attenuating Fourier components that do not exceed the amplitude
of the active silence by a predetermined factor. 

Noise-gating to remove $N_{a}$ can be adapted in a straightforward way
to image sequences, by segmenting (and Fourier transforming) in two
($x,y$) or three ($x,y,t$) dimensions rather than one. However,
this adaptation is not readily applicable to many image applications,
because uniform additive noise is often not the dominant contaminant
of image data. In a large class of images, $N_{s}$, rather than $N_{a}$,
dominates the noise field in the image: $N_{a}$ and $N_{other}$
can be neglected. Since $N_{s}$ depends on the value of $Im_{0}$,
filtering based on a simple threshold of component amplitude is not
sufficient as it is for $N_{a}$.

\paragraph{Estimating noise level}

Shot noise arises from the Poisson-distribution statistics of counting
discrete quanta -- photons, photoelectrons, or other quantized elements
that depend on the detection technology. $N_{s}$ is then a random
variable, sampled once per pixel, whose value depends also on the
local value of $Im_{0}$. When the number of quanta is high, the Poisson
distribution is well approximated by a normal (Gaussian) distribution, and
an estimate of the shot noise can be written directly: 
\begin{equation}
N_{s}(x,y,t)\approx\alpha G(x,y,t)\sqrt{Im_{0}(x,y,t)},\label{eq:shotnoise equation}
\end{equation}
where $\alpha$ is an instrument-dependent constant, and $G(x,y,t)$ is
a random variable with a fixed Gaussian distribution of mean 0 and
variance 1. Equation \ref{eq:shotnoise equation} is particularly
useful because it divides shot noise into three components, two of
which can be characterized well. The $\alpha$ coefficient is a
constant of the instrument, and can be reconstructed from flat-field
images or directly from image data, and $G(x,y,t)$ is a standard tool
of statistical analysis. By Fourier transforming Equation
\ref{eq:shotnoise equation} it is possible to estimate the spectral
amplitude of the noise in the image:
\begin{equation}
N_{s}'(k_{x},k_{y},\omega)\approx\alpha G'(k_{x},k_{y},\omega)\otimes\mathbb{F}\left(\sqrt{Im_{0}(x,y,t)}\right).\label{eq:shotnoise spectral equation}
\end{equation}
But $G'(k_{x},k_{y},\omega)$ is constant across Fourier space, since
$G(x,y,t)$ is a normal random variable; and for a very broad class
of scenes the zero-frequency component of the Fourier transform dominates
the spectrum and the $F(\sqrt{Im(x,y,t)})$ term can be treated as
a delta function. Hence we arrive at the approximation:
\begin{equation}
\left|N'_{s}(k_{x},k_{y},\omega)\right|\approx\beta(k_{x},k_{y},\omega)\sum_{x,y,t}\sqrt{Im(x,y,t)},\label{eq:noise-estimate}
\end{equation}
where $\beta$ is a constant-across-images spectrum that is characteristic
of the instrument that acquired the image, and we've used the fact
that the sum of the noise term over many pixels is approximately zero,
to replace $Im_{0}$ with $Im$ under the radical. 

Equation \ref{eq:noise-estimate} is useful because it estimates the
noise amplitude in a given image or image subregion, provided that
$\beta$ can be determined. In principle, $\beta$ can be determined\emph{
a priori} from the absolute sensitivity of the instrument, but it
is also accessible via \emph{a posteriori }analysis of the data themselves.
This is accomplished by breaking a full dataset $Im(x,y,t)$ into
multiple small samples $Im_{i}$ and searching for a minimum scaled
spectrum for each one. Fourier transforming Equation \ref{eq:image plus noise}
(with $N_{a}$ and $N_{other}$ neglected) and substituting Equation
\ref{eq:noise-estimate} yields
\begin{equation}
Im'_{i}(k_{x},k_{y},\omega)\approx Im'_{0,i}(k_{x},k_{y},\omega)+\beta\sum_{x,y,t}\sqrt{Im_{i}(x,y,t)}\label{eq:spectrum}
\end{equation}
for each subimage index $i$. Solving for $\beta$ gives
\begin{equation}
\beta_{i}(k_{x},k_{y},\omega)\approx\frac{\left|Im_{i}'(k_{x},k_{y},\omega)-Im_{0,i}'(k_{x},k_{y},\omega)\right|}{\overline{Im_{i}}},\label{eq:beta-formula}
\end{equation}
where the difference in the numerator is still just the (unknown)
Fourier spectrum of the shot noise, and the bar over $Im_{i}$ indicates
summing the square root of each (positive-definite) pixel value. Across
a large population of samples, the estimates of $\beta_{i}$ at a
given location in Fourier space will vary from a minimum where the
noise spectrum sample at that particular point is near zero, to a
maximum where the local sampled value of the shot noise is much larger
than the corresponding Fourier component $Im_{0}'.$ But structured
images and image sequences containing coherent features are dominated
by a few sparse Fourier components where $Im'_{0}(k_{x},k_{y},\omega)\approx Im'(k_{x},k_{y},\omega).$
Because the shot noise is a random variable, its Fourier amplitude
is more nearly constant throughout the space, and most Fourier components
are instead dominated by the noise: $\left|Im'_{0}\right|\ll\left|Im'(k_{x},k_{y},\omega)\right|$.
Because the latter is the more common case, the median value of $\beta_{i}$
across many image samples is a good estimator of the noise spectrum,
and we can take
\begin{equation}
\beta_{approx}(k_{x},k_{y},\omega)=median_{i}\left(\frac{\left|Im_{i}'(k_{x},k_{y},\omega)\right|}{\overline{Im_{i}}}\right),\label{eq:beta-calc}
\end{equation}
which depends only on the statistics across image subsamples of the
Fourier spectrum in the original data set. The approximation in Equation
\ref{eq:beta-calc} requires that a significant fraction of Fourier
space be noise dominated: at least half of all samples. This is typically
the case in image sequences that have direct visual evidence of shot
noise, but for images that are more highly structured the median could
be replaced with a lower percentile value. The calculated value of
$\beta_{approx}$ allows estimation of the noise level across \emph{all}
regions of an image sequence dominated by conventional shot noise,
per Equation \ref{eq:noise-estimate}.

Of course, for image-\emph{independent }additive noise, one can estimate
the noise level with a simpler calculation of a constant level across
image segments: 
\begin{equation}
\left|N'_{i,a}(k_{x},k_{y},\omega)\right|=median_{j}\left(Im'_{j}(k_{x},k_{y},\omega)\right),\label{eq:additive-noise}
\end{equation}
where (as in Equation \ref{eq:beta-calc}) one may replace the median
with a lower percentile across image samples in the case that noise
is low or the image is very highly structured.

\paragraph{Filtering}

Having produced a noise model, one can generate an adaptive filter
tuned to the estimated noise spectrum in each sample:
\begin{equation}
\check{Im'_{i}}(k_{x},k_{y},\omega)\equiv Im_{i}'(k_{x},k_{y},k_{\omega})F'_{i}(k_{x},k_{y},k_{z}),\label{eq:noise-gate}
\end{equation}
where $F'_{i}$ is a filter function of some sort. For the simplest
processing, $F_{i}$ is the gating function: 
\begin{equation}
F'_{i,gate}(k_{x},k_{y},\omega)\equiv\left\{ {\begin{array}{cc}
0 & \mathrm{if}\,Im'_{i}(k_{x},k_{y},\omega)<T_{i}(k_{x},k_{y},\omega)\\
1 & \mathrm{otherwise}
\end{array}}\right\} ,\label{eq:gate}
\end{equation}
where $T_{i}$ is a threshold function based on the noise levels from
Equations \ref{eq:noise-estimate} or \ref{eq:additive-noise}. The
Wiener filter 
\begin{equation}
F'_{i,wiener}(k_{x},k_{y},\omega)\equiv\frac{Im'_{i}(k_{x},k_{y},\omega)/T_{i}(k_{x},k_{y},\omega)}{1+Im'_{i}(k_{x},k_{y},\omega)/T_{i}(k_{x},k_{y},\omega)}\label{eq:wiener}
\end{equation}
rolls off filter response more gradually but may admit more noise
at a given threshold level.

In practice, the threshold can be defined using an \emph{ad hoc }factor
$\gamma$ to bias the filtering between the preference to preserve
the most signal possible or to reject the most noise possible: 
\begin{equation}
T_{i}(k_{x},k_{y},\omega)\equiv\gamma N'_{i}(k_{x},k_{y},\omega)\label{eq:thresh-definition}
\end{equation}
for whichever noise model is most appropriate. For the solar applications
in Section \ref{sec:Results}, $\gamma=3$ provides a good balance
between recognizable features and noise reduction.

\paragraph{Apodization \& reconstruction}

As with all Fourier methods, the noise gating described here requires
careful apodization of the $Im_{i}$ image segments. Apodization brings
the edges of each $Im_{i}$ smoothly to zero in a way that minimizes
edge effects on the Fourier spectrum. It is accomplished by multiplying
each image sample by a windowing function $w(x,y,z)$:
\begin{equation}
Im_{i}(x,y,z)=w(x,y,z)Im_{i,raw}(x,y,z)\label{eq:windowing-operation}
\end{equation}
where $Im_{i,raw}$ represents a small sample of data in an image
sequence and $Im_{i}(x,y,z)$ is the image function treated above.
Multiplying by a windowing function convolves the Fourier spectrum:
\begin{equation}
Im'_{i}(k_{x},k_{y},\omega)=Im'_{i,raw}(k_{x},y_{y},\omega)\otimes w'(k_{x},k_{y},\omega).\label{eq:window-convolution}
\end{equation}
The optimal windowing function is the Hanning window: 
\begin{equation}
w_{h}(x,y,t)=sin^{2}((x+0.5)\frac{\pi}{n_{x}})sin^{2}((y+0.5)\frac{\pi}{n_{y}})sin^{2}((t+0.5)\frac{\pi}{n_{t}},\label{eq:window}
\end{equation}
where $n_{x}$, $n_{y}$, and $n_{t}$ are the size (in pixels) of
each dimension of $Im_{i}$. The Hanning window has Fourier power
only in the zero and first nonzero frequency in each axis, and therefore
minimally spreads the spectrum of the apodized image. 

Hanning windows are also ideal for smooth image reconstruction, because
$sin^{2}(\alpha)+cos^{2}(\alpha)=1$: by overlapping adjacent image
samples by 50\%, one can simply sum adjacent apodized samples to reconstruct
the full original (pre-apodization) image sequence. This approach
is, in fact, used in one dimension in popular Fourier audio compression
schemes such as MP3. However, in this application the summation properties
of the apodization function are lost because of the noise gating itself.  
The Fourier-space product between the filter and the transformed, apodized
image block is equivalent to a convolution in real space, which damages the
profile of the window function.

There are various rigorous techniques for building windowing functions
that avoid this problem. The most simple is applying a double-Hanning
window: set 
\begin{equation}
\check{Im_{i,final}}(x,y,t)\equiv w_{h}\check{Im_{i}}=w_{h}G\left(w_{h}Im_{i,raw}\right),\label{eq:double-gating}
\end{equation}
where the $G$ operator represents the entire noise-gating sequence
described above. If the gating step in Fourier space were the no-op,
this would result in a simple windowing by $sin^{4}$. The original
image can be reconstructed by oversampling the width of $Im_{i}$
by a factor of 4 in each axis, and summing all the resulting image
segments. This is because
\begin{equation}
\sum_{j=0}^{3}sin^{4}(x+\frac{j\pi}{4})=1.5,\label{eq:reconstruction-sum}
\end{equation}
for all $x$. The oversampling and additional windowing attenuate
any edge artifacts caused by the filtered gating function.

\paragraph{Implementation}

I implemented this calculation in the Perl/PDL language and executed
it on several scientific data sets, notably EUV images of the Sun.
Data sets were subsampled to $12\times12\times12$ pixel sequences
staggered every 3 pixels in space and time, with the double-Hanning
apodization described above. The code is available as part of the
``solarpdl-tools'' distribution at GitHub.com (in the file 
"image/noise-gate-batch.pdl"). It is also attached to
the digital version of this article. Several small adaptations have
been made in application. For example, rather than the full complex
discrete Fourier transform the code uses the real discrete
Fourier transform or, equivalently, the discrete cosine transform.
Several "wrapper" routines implement multitasking and/or stream
processing of a potentially semi-infinite set of images.

I implemented the code in Perl/PDL because of that environment's superior handling of
high-dimensional objects, ready library of scientific application modules,
simple process control for multitasking, and ease of adaptation to compiled code.
The existing code makes heavy use of the PDL vectorization engine.  This is 
convenient for prototyping, but (as with all vectorized languages, including 
Numeric Python and IDL) it is pessimal for cache maintenance.  Hence, the code
almost certainly runs an order of magnitude slower than an optimized application
written in a fully compiled language such as C or FORTRAN. 

Gating with the existing code is feasible even for large image sequences (such as
full-frame SDO/AIA images) on current hardware.  Stream processing AIA images with
block size 12x12x12 and 4x oversampling requires 30-90 seconds per frame on typical 
recent (2016) hardware, with the longer time representing a laptop computer and
the shorter a rackmount "CPU server" with high speed internal bus.  A cache-optimized
implementation might be expected to run 10x faster, and a GPU-optimized or similar highly
parallel implementation might run 10x faster still.

\section{\label{sec:Results}Results}

Here I present results of applying the noise gating algorithm described
in §\ref{sec:Background-=000026-Technique} to images and image sequences.
Noise gating works best on image sequences because of the additional
feature coherence afforded by the time dimension, and examples in
Sections \ref{sub:Shot-noise:-Coronal} and \ref{sub:Additive-noise:-solar}
show noise reduction in different types of solar image sequences.
The technique is also applicable to individual images, and Sections
\ref{sub:Single-image-shot-noise} and \ref{sub:A-conventional-photograph:}
demonstrate performance in those domains.

\subsection{\label{sub:Shot-noise:-Coronal}Shot noise: Coronal image sequences
from SDO/AIA}

Extreme-ultraviolet images of the solar corona have revolutionized
solar physics since the 1990s. These images are collected in short
spectral lines emitted by very hot, multiply-ionized trace elements
(such as iron). The Atmospheric Imaging Assembly on board the \emph{Solar
Dynamics Observatory} (SDO/AIA; \citealt{lemen_atmospheric_2011})
produces $4096\times4096$ pixel EUV images of the Sun in several
spectral lines, once every 12 seconds. This high cadence is intended
to capture evolution and dynamics of both large- and small-scale features.
At the smallest scales visible in AIA images (up to a few seconds
of arc), most features are dominated by shot noise in the original
images. This makes them susceptible to improvement by scaled noise
gating as described in §\ref{sec:Background-=000026-Technique}.

\begin{figure}

\begin{centering}
\includegraphics[width=6in]{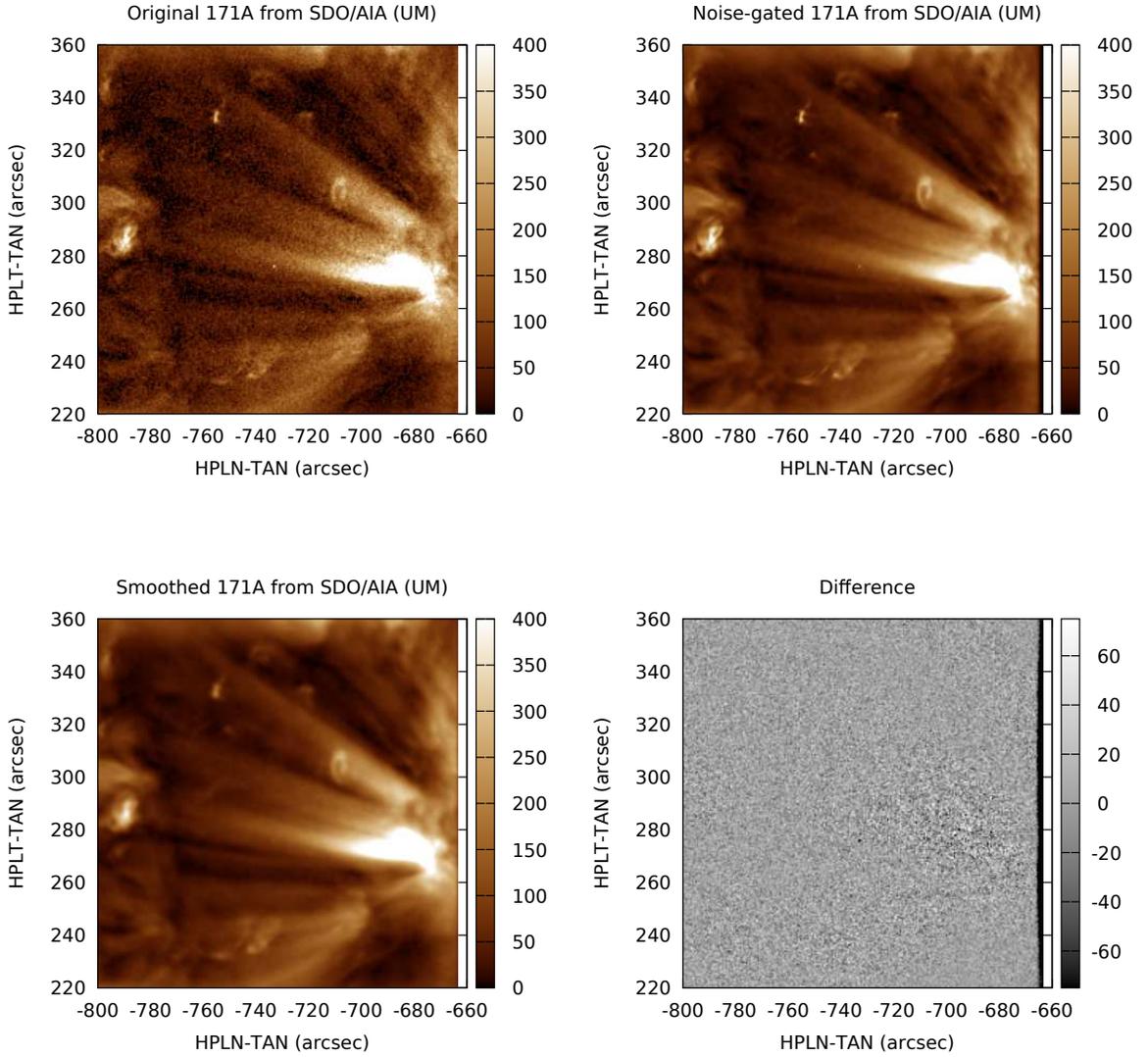}
\par\end{centering}

\caption{\label{fig:AIA-demo}Noise gating significantly improves \emph{SDO}/AIA
EUV images, which are limited by conventional shot noise, as seen
in this unsharp-masked closeup of a low-latitude coronal plume. Clockwise
from top left: Original L1 image is strongly noise limited; noise-gated
image preserves resolution while deeply suppressing the shot noise
by more than $10\times$; residual differences between the original
and gated image show no overall structure; Gaussian-smoothed image
achieves comparable suppression, but with $5\times$ degradation in
both spatial and temporal resolution. See also the movie attached
to the digital version of this article, which shows the improvement 
still more dramatically.}

\end{figure}

Figure \ref{fig:AIA-demo} shows the limitations imposed by shot noise
on the AIA images.  It is available as a still frame and also (in the
digital version of this article) as a movie.  The differences are apparent
in the still figure, but are yet more visually striking in the movie.  
The upper-left panel is
an image of a mid-latitude coronal plume, acquired on 2016 July 4 in
that instrument's 17.1 nm passband. It is a subfield of the SDO/AIA "Level 1"
data set that is available via the virtual solar observatory (http://vso.stanford.edu)
or the mission's primary data distribution center (http://jsoc.stanford.edu). 
The image has been ``unsharp
masked'': a smooth background, made with the \emph{minsmooth} operator
(\citet{deforest_fading_2016}) has been subtracted from it.  The smooth
background is made from a local-minimum estimator with an aperture
diameter of 30 arcsec, so that the final image is positive definite
but contains only features with a spatial scale (in at least one
direction) less than 30 arcsec.  The image is easily seen to be
affected by shot noise from the quantization of the EUV flux coming
from the feature.

The lower-left panel of Figure \ref{fig:AIA-demo} has been smoothed
with a 5-pixel full-width spherical Gaussian kernel in both space and
time.  The shot noise has been reduced by a factor just over 10, at
the cost of a 5$\times$ reduction in both spatial and temporal
resolution (i.e. a reduction of the dataset to under 1\% of its
original information content).

The upper-right panel of Figure \ref{fig:AIA-demo} shows the effect of
noise-gating as described in Section
\ref{sec:Background-=000026-Technique}, using a 12x12x12 pixel
subregions with double-Hanning windowing, direct threshold gating of
Fourier components and a $\gamma$ factor of 3.  The noise is reduced
by approximately the same factor as the smoothed image, but high
spatial and temporal resolution are preserved.  In particular, small
bright points and ejecta, many of which are not directly noticeable in
the original data, are preserved in the noise-gated sequence even
though they are lost in the original and/or in the smoothed sequence.

Moreover, myriad small bright ejecta in the plume itself are visible in
the gated sequence, traveling from the core near (-680,270) leftward
along the bright fibrils of the plume structure.  At least some of
these ejecta, once spotted, can be identified in the original data,
but are lost in the smoothed data.  Others can be spotted with frame-to-frame
stepping of the gated sequence, but are too far below the noise floor
of the original data to spot by eye.

The lower-right panel shows the residual (difference) between the
original image and the noise-gated image in the top two panels.  While
the noise level varies as expected, being stronger in the brighter portions
of the image, it contains no visually identifiable image structure.
The normalized correlation coefficient between the difference image and
the noise-gated image fluctuates frame-to-frame, is 0 on average, and
has a value well under 1\% for individual frames.

It is worth noting that the residual (difference) images do \emph{not}
show signs of the faint ejecta along the plume, indicating that they
are present in the original data although they are difficult to discern
against the noise floor.

\subsection{\label{sub:Additive-noise:-solar}Additive noise: solar magnetogram
sequences from SDO/HMI}

Solar magnetograms are image data products that represent the amount
of magnetic flux penetrating the surface of the Sun. The Helioseismic
and Magnetic Imager on board the \emph{Solar Dynamics Observatory}
(SDO/HMI; \citealt{scherrer_helioseismic_2012}) downlinks polarized
images in multiple narrow spectral bands (``filtergrams''), spanning
a single spectral absorption line. These images are assembled \emph{post
facto} into a data product that represents the line-of-sight magnetic
field on the surface of the Sun, exploiting the Zeeman effect. The
noise spectrum of resulting ``magnetogram'' images is dominated
by shot noise in the individual filtergrams, but is not dependent
on the local value of the magnetogram data product. Hence it can be
modeled as additive noise on the data product itself. The method outlined
in §\ref{sec:Background-=000026-Technique} is therefore applicable,
using Equation \ref{eq:additive-noise} to define a threshold spatial
spectrum.

\begin{figure}
\begin{centering}
\includegraphics[width=5in]{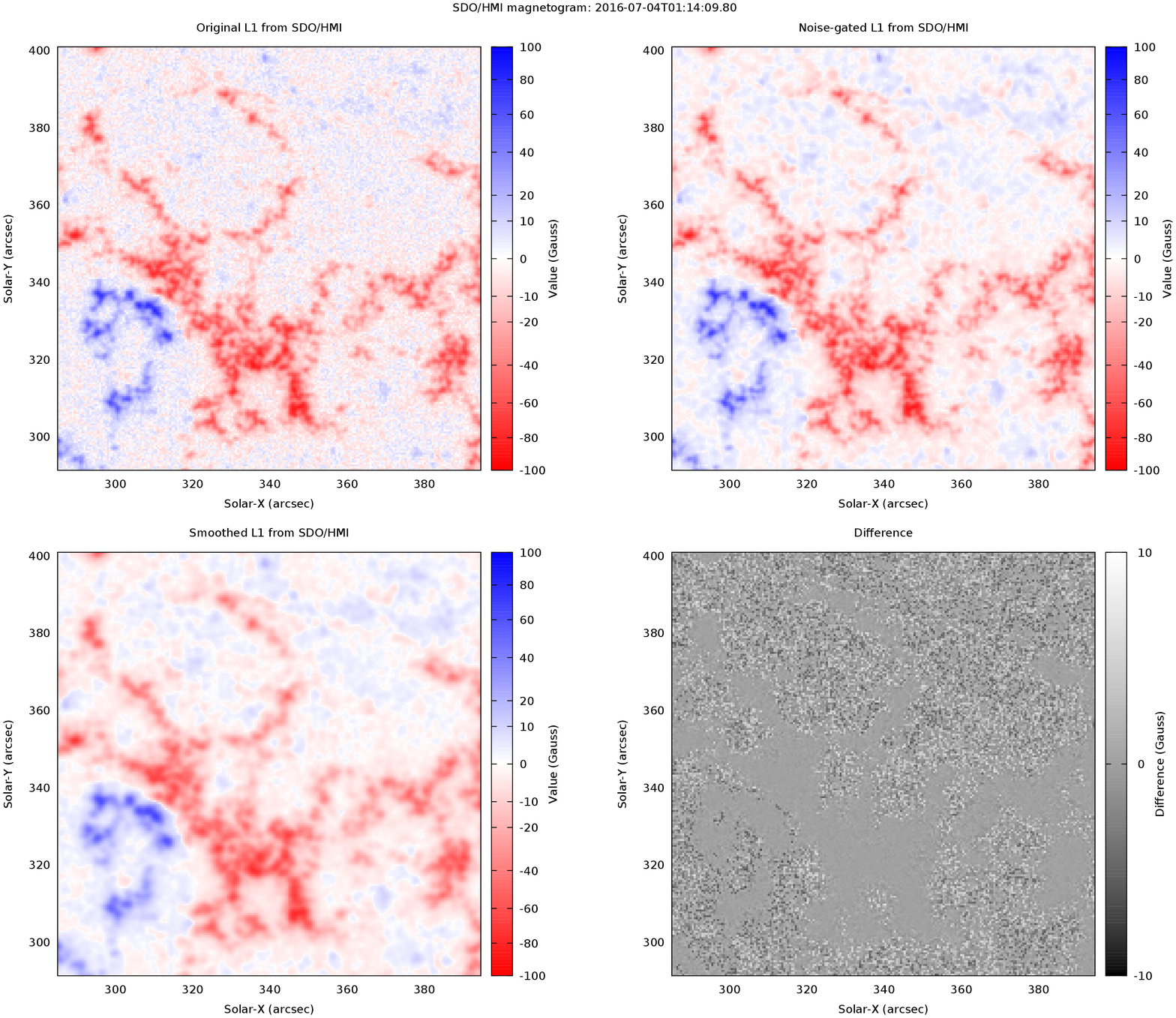}
\par\end{centering}

\caption{\label{fig:Fixed-threshold-noise-gating}Noise gating improves \emph{SDO}/HMI
magnetograms, which are are limited by image-independent additive
noise. Clockwise from top left: close-up of an original HMI 45-second
magnetogram; the same magnetogram, gated as described in the text;
difference image between the gated and original magnetograms; and
a smoothed magnetogram with similar noise spectrum to the gated image.
Gating reduces noise in the quiet regions of the image, without incurring
the spatial or temporal resolution ``hit'' of smoothing where the
signal is strong. (This figure is also available as a movie in the digital
version of this article.)}
\end{figure}

Figure \ref{fig:Fixed-threshold-noise-gating} shows processing of
an example image sequence in extreme close-up ($240\times240$ pixels
compared to the native $4096\times4096$) of a sequence of 128 magnetograms
from SDO/HMI. These data are available from the same locations as SDO/AIA
data (Section \ref{sub:Shot-noise:-Coronal}, above).  The fixed spectrum 
threshold was set as $4\times$ the
25 percentile value of each spectral component across all $8\times8\times8$
data samples in the magnetogram sequence ($5\times10^{5}$ samples
in all).

\begin{figure}
\begin{centering}
\includegraphics[width=5in]{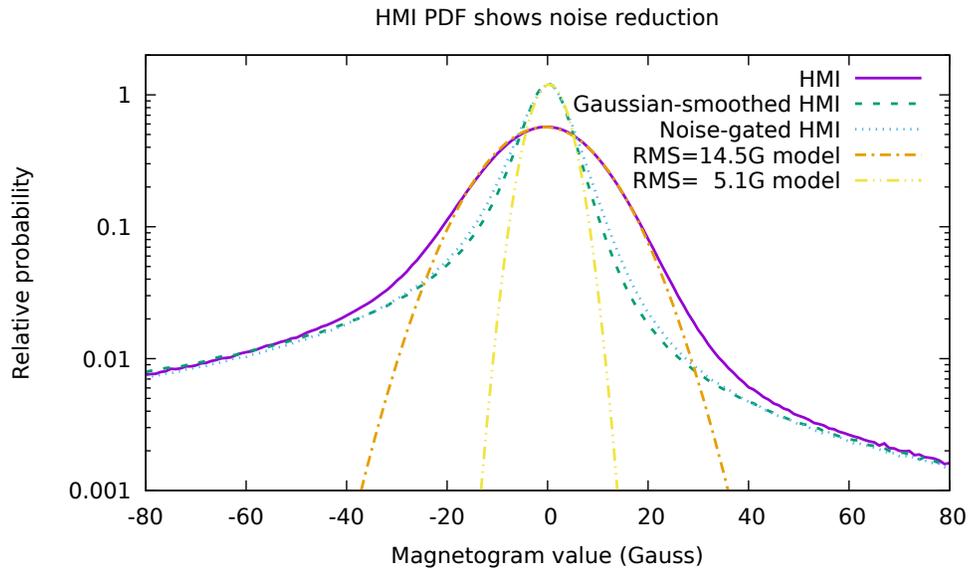}
\par\end{centering}

\caption{\label{fig:Probability-distribution-functio}Probability distribution
function of the image value in Figure \ref{fig:Fixed-threshold-noise-gating}
shows the effects of noise and its reduction. Strong magnetic features
form broad tails in the PDF. Additive shot noise forms a Gaussian
core (inverted parabola on this semilog plot) with $\sigma=$14.5G.
Both noise-gating and smoothing reduce the the Gaussian core to $\sigma=6.0$G. }
\end{figure}

Because magnetograms are subject to additive noise and have most
values near zero, the noise level can be seen through direct visual
inspection of the probability distribution function
(e.g. \citealt{schrijver_sustaining_1997}) as shown in Figure
\ref{fig:Probability-distribution-functio}.  The additive noise in the
magnetograms is well approximated as a normal (Gaussian) additive
source with $\sigma$=14.5 G, which appears on the semilog plot as an
inverted parabola, centered on zero, as the core of the magnetogram's
probability distribution function.  The width of the parabola
represents the noise level.  Smoothing the original with a spherical
Gaussian kernel with full width of $3$ pixels reduces $\sigma$ of this
distribution core to 5.1G, at the cost of reducing the effective
number of datapoints by a factor of roughly 10. Noise-gating removes
noise from regions where it would dominate the magnetic signal, both
revealing weak magnetic features and preserving the full instrument
resolution in strong features.

The difference image at lower right in Figure
\ref{fig:Fixed-threshold-noise-gating} shows the effectiveness of the
technique at preserving strong magnetic features: the portions of the
image that are dominated by strong magnetic signal are preserved
exactly as in the original data. Weak-field regions have high spatial
frequencies removed, preserving only those elements of image structure
that rise above the noise ``floor''.  While the photospheric magnetic
field evolves relatively slowly compared to the overlying corona, at
the very smallest scales flux emergence and/or cancellation is visible
on timescales of 3-5 frames.  As with Figure \ref{fig:AIA-demo}, the
attached digital movie is yet more visually striking than the still frame.

\begin{figure}

\begin{centering}
\includegraphics[width=5in]{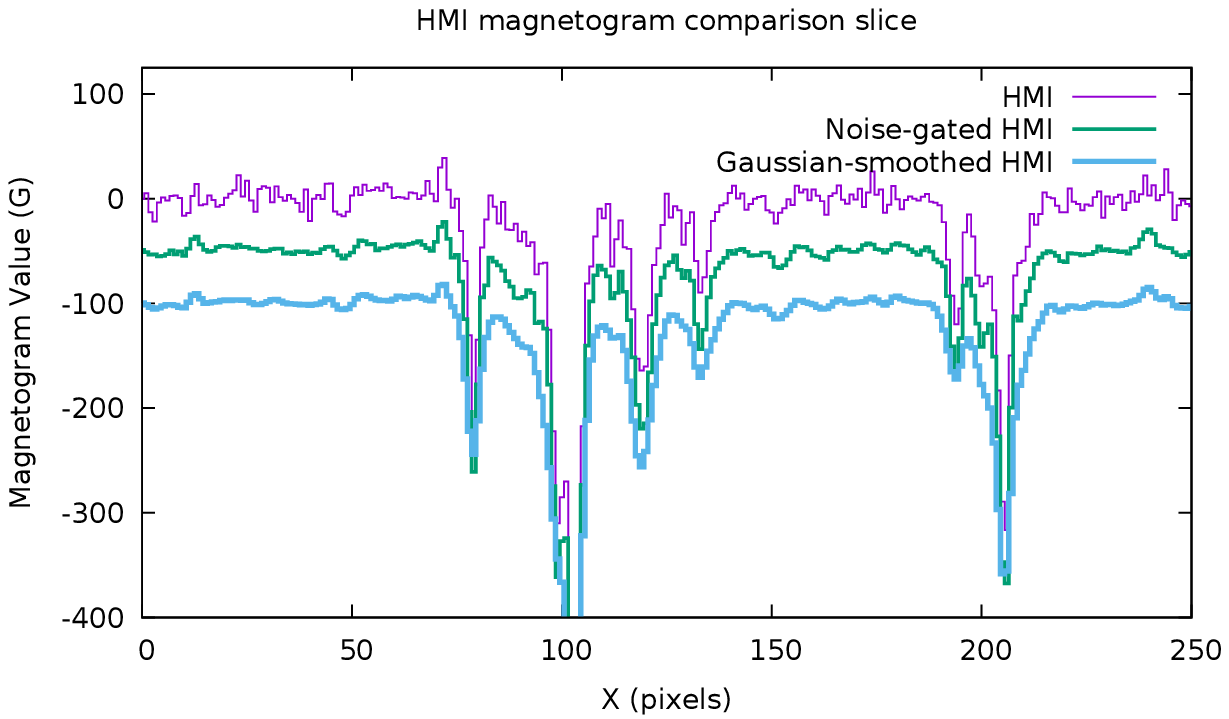}
\par\end{centering}

\caption{\label{fig:Slices-through-the}Single slice through the HMI image
shown in Figure \ref{fig:Fixed-threshold-noise-gating} reveals that
individual features' ``photometry'' and detail are preserved despite
rejection of high-spatial-frequency noise. Traces are offset vertically
by 50G for clarity. For comparison to Figure \ref{fig:Fixed-threshold-noise-gating},
HMI pixels subtend 0.5 arcsec.}
\end{figure}

Overall ``photometry'' and feature preservation is further
demonstrated in Figure \ref{fig:Slices-through-the}, which is a plot
of the results of noise-gating and Gaussian smoothing on a single
horizontal slice of the image in Figure
\ref{fig:Fixed-threshold-noise-gating}. The gated data and smoothed
data have roughly equal attenuations of the high spatial frequency
``noise'' -- but the noise-gated trace preserves strong mixed-polarity
flux and magnetic features at the instrument resolution, while the
smoothed line does not.

\subsection{\label{sub:Single-image-shot-noise}Single-image shot noise gating:
EUV images from NIXT}

\begin{figure}
\begin{centering}
\includegraphics[width=6in]{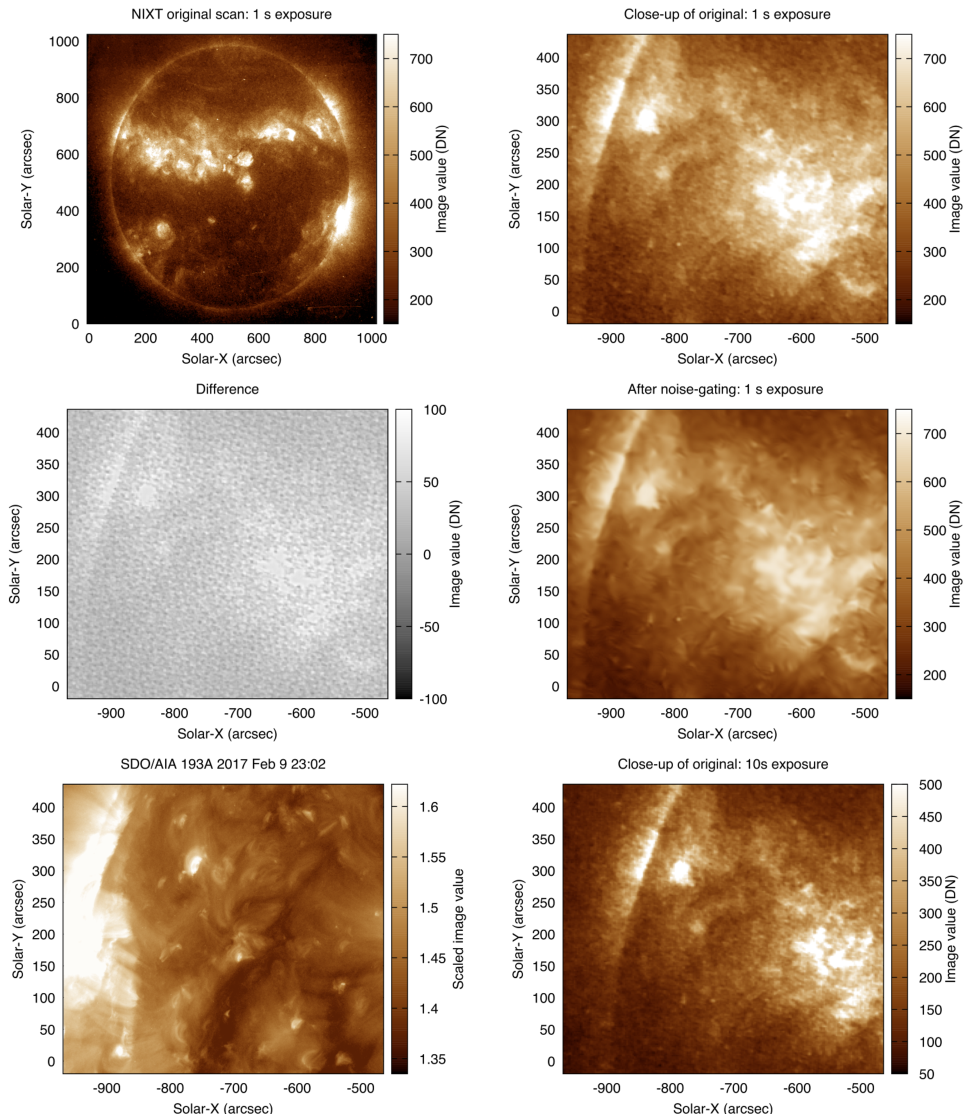}
\par\end{centering}

\caption{\label{fig:NIXT}Enhancement of a 1 s exposure NIXT image of the Sun
in the 193Å Fe emission line, from a sounding rocket flight on 1989
Sep 11, reveals fine scale coronal loops not visible in the original
image. Clockwise from top left: full-Sun image; close-up of an active
region at upper left; noise-gated image of the active region reveals
small loop structures in the nearby corona; longer exposure (un-enhanced)
from NIXT reveals the same loops; modern 193Å image from \emph{SDO}/AIA,
scaled to match the NIXT film response, reveals similar small, faint
structure in the corona; difference image between the processed and
unprocessed NIXT reveals high-spatial-frequency components removed
from the image.}

\end{figure}

Although the noise-gating works best on image sequences, it is also
suitable for enhancement of still-frame images. The Normal Incidence
X-Ray Telescope (NIXT) was a sounding rocket payload flown in the
late 1980s that, together with its sister rocket the MSSTA, prototyped
all modern EUV solar telescopes (e.g., \citealt{golub_sub-arcsecond_1990},
Walker et al. \citeyear{walker_soft_1988}). These early sounding
rockets used photographic film and were strongly limited by both film
grain and photon shot noise throughout most of the field of view,
though NIXT did achieve full sub-arcsec resolution in bright features.
Applying the noise-gating to NIXT images reveals more fine coronal
structure than is apparent in the images themselves. Figure \ref{fig:NIXT}
shows the enhancement. An active region in the northeast (upper left)
quadrant of the Sun is dominated by film grain (upper right panel).
Noise-gating reduces the effect of the film grain (center left), revealing
myriad faint fine-scale loops (center right; several are visible between
X\textasciitilde{} -700 , y \textasciitilde{}100). The features could
in principle be artifacts, but they are visible in an unprocessed
longer exposure from the same rocket flight (lower right panel), verifying
that they are real solar features being exposed by noise reduction. 

The processed NIXT image has a different ``texture'' than many modern
digital EUV images acquired in this spectral line, but that is attributable
to remnant film grain and to the highly nonlinear response of photographic
film to solar fluence. Images from the \emph{SDO}/AIA 193Å channel
reveal similar fine-scale structure in the corona (lower left panel)
when the contrast is adjusted appropriately -- in this case by taking
the 12$^{th}$ root of the reported image values (raising them to
the power 0.083).

The NIXT images are a challenging target for this algorithm: they
are individual stills with different exposure characteristics, limiting
analysis to 2-D rather than 3-D; the detector response is highly nonlinear
(the best-match AIA contrast profile scales as the twelfth root of
the instrument-reported radiance); and the images are affected by
a mixture of shot noise and film grain. Nevertheless, these historical
images are noticeably improved and reveal features in the corona that
were present, but invisible, before processing.

\subsection{\label{sub:A-conventional-photograph:}A conventional photograph:
the standard Lena test image }

\begin{figure}
\begin{centering}
\includegraphics[width=5in]{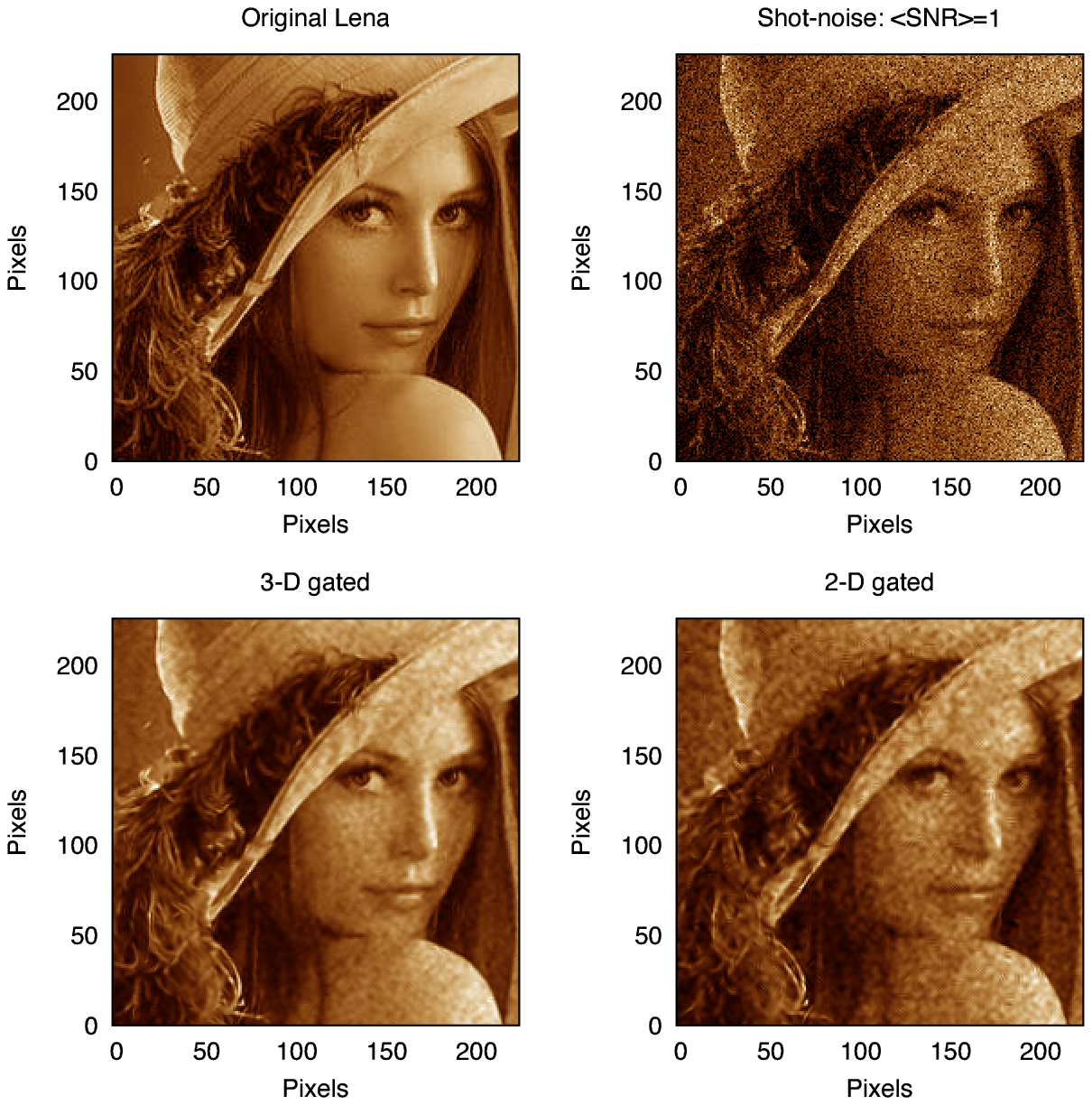}
\par\end{centering}

\caption{\label{fig:lena-figure}Shot-noise degradation and restoration of
the standard ``Lena'' image shows the advantage of 3-D \emph{sequence}
processing for additional coherence. Clockwise from top left: original
``Lena'' image; degraded with shot noise: $\left<SNR\right>=1$;
restored with 2-D noise gating: $\left<SNR\right>=6$; restored with
3-D noise gating: $\left<SNR\right>$=14.}
\end{figure}

I applied noise-gating not only to astrophysical images but, for reference,
to ``Lena'', an industry-standard image adopted by the image processing
community for over four decades. Figure \ref{fig:lena-figure} shows
the effect of noise-gating on a noise-degraded version of the ``Lena''
image. 

To highlight the difference between 2-D and 3-D processing, I simulated
a slowly varying scene using the Lena image. I duplicated the image
16 times with a slow parametric shift of 0.5\textdegree{} rotation
and 0.5 pixel displacement per frame, with the middle frame (no. 8)
in its original position. I then added a different instance of the
shot noise model to each copy. Noise-gating in 2-D improved the SNR
by a factor of 6, and in 3-D by a factor of 14. Both applications
recovered subtle and small image features that are apparently lost
in the low-SNR degraded copy.

The additional noise reduction for 3-D processing arises because the
Fourier method exploits feature coherence across both time and space.
It requires only that the moving/evolving object have a discrete Fourier
spectrum that is distinguishable in amplitude from the background
noise field, without regard for the shape of the feature or the particulars
of its evolution across image frames.

While the noise gating output does not have the visual cleanliness
of many of the more cosmetic denoisers (for a good overview, see \citealt{buades_review_2005}),
it does have direct roots in photometric noise reduction, and (as
demonstrated in Section \ref{sub:Additive-noise:-solar}) preserves
photometry of small features. The remaining shot noise is commensurate
with the actual sensitivity obtained by conventional averaging over
the same size as the neighborhoods used for local Fourier transformation.
In particular, the shot noise in the lower left panel of \ref{fig:lena-figure}
is reduced by $14\times$ by the 3-D noise-gating, i.e. within a factor
of two of the ideal photometric improvement obtained simply
by averaging the noisy value across each $8\times8\times8$ pixel
region of the synthetic image sequence. The improvement over smoothing
arises from preservation of fine detail in places where the image
contrast is sufficiently high. In particular the sharp edges of the
hat's brim and feather boa ribbon are preserved at full spatial resolution
despite the low SNR in the degraded source image.

\section{\label{sec:Discussion-=000026-Conclusions}Discussion }

Normal-distribution random additive noise with fixed or variable characteristics
(generally photon shot noise, though there are other sources as well)
affects nearly all scientific images. The most commonly used approaches
to overcome it are smoothing of various sorts, and better approaches
are available even without specific reference to the content of the
data. 3-D noise gating can greatly improve data affected by this type
of noise, by discriminating more precisely between likely noise and
likely signal than can conventional content-blind methods. 

Convolutional smoothing \textendash{} in space, time, or both \textendash{}
sacrifices resolution for noise mainly because the method is not signal-aware.
Model fitting is generally extremely content-aware and therefore requires
prior knowledge of the particular subject being recorded. Noise gating
is a signal recovery ``happy medium'': it is one of a large class
of adaptive filters that respond to the characteristics of a dataset
to improve discrimination between signal and noise, without prior
knowledge of the signal itself. 

Noise gating works because the prime discriminant between ``signal''
and ``noise'' is structural coherence, which yields peaks in the
Fourier spectrum of each neighborhood in the data. Because discernible
features have spatial and temporal extent, they are generally concentrated
by the Fourier transform into a smaller region in Fourier space than
in the direct space of independent variables, which affords easier
discrimination between the feature and the backround ``noise''. 

For a broad range of applications, noise-gating is therefore a large
improvement over traditional noise-reduction techniques. By discriminating
noise from signal in the Fourier domain 3-D noise gating greatly reduces
noise level without significant loss of spatial or temporal resolution.

There is further room for improvement. I've presented here a simple
adaptive filter for rejecting variable or fixed additive noise, whose
character depends on local signal brightness but not on location in
the data. Obvious extensions include filters that are aware of instrument
focal variation or vignetting, of multiple (compound) noise sources
such as shot noise mixed with non-negligible read noise, or of multiplicative
or non-normally-distributed noise sources. Although they are not considered
here, there are also obvious generalizations to color or hyperspectral
images, which might take advantage of correlation between different
wavelength channels or filter images in indirect color spaces such
as the hue-saturation-value (HSV) system, rather than in direct
wavelength-channel intensity spaces such as red-green-blue (RGB).

Noise-gating does have limitations. In particular, either form I've
presented here (strict gating and Wiener filtering) either obliterates
or strongly attenuates any weak signal that cannot be discriminated
from noise using the neighborhood Fourier amplitudes. This highlights
and reveals features and signals that may not have been apparent before
processing, but prevents subsequent lock-in or related measurements
that could ``drill'' well below the noise floor, acting on the original
data. Therefore noise-gating should be applied as the \emph{last}
step in a multi-step image reduction pipeline. In this respect, its
limitations are similar to those of convolutional smoothing and related
denoising techniques, but with the significant advantage of retaining
full spatiotemporal resolution in the cleaned images. 

I have demonstrated noise-gating primarily in solar applications, in part
because data-rich image sequences are so readily available and so important
to that subfield of astrophysics.  But the technique is applicable to other
astrophysical targets and techniques, including wide-field imaging of the solar 
wind itself, stellar-coronagraph exoplanet imaging, nebula evolution studies,
time-lapse studies of supernova remnants such as Eta Carina, and time-lapse 
orbital analysis of globular clusters.  The technique has broad applicability
to subjects outside of astronomy and astrophysics, though I have focused on 
astrophysical applications in this article.

\section{\label{sec:Conclusions}Conclusions}

I have introduced and demonstrated a novel denoising technique, 3-D
noise gating, that is suitable for use in scientific grade astrophysical
images. The technique removes shot noise from image sequences while
retaining photometry of coherent image features at the full resolution
of the original data. Noise-gating can also be applied in 2-D for
still images, and has been known for some time in that context although
it appears not to have been adopted by the astrophysical community.
Noise-gating is useful in part because it relies primarily on feature
coherence as a discriminant, without regard to any particular shape,
motion profile, or evolutionary characteristic of the desired signal.
In that regard it is a powerful general-purpose technique for data
improvement. In the applications I have demonstrated, it reduces shot
noise by a factor of order 10, and is therefore equivalent to an increase
in exposure time between 10$\times$ and $100\mbox{\ensuremath{\times}}$
for revealing faint features in noise-limited astrophysical image
sequences. The cost is that very faint features below the reduced
noise floor, even those that could in principle be recovered from
more targeted techniques, are obliterated. The algorithm has been
implemented in Perl Data Language with C libraries, is easily
encapsulated for general-purpose use, and is readily portable to other
computing environments.

\acknowledgements{This work was funded by NASA grants NNX-16AF98G and
  NNX-16AG98G.  Thanks are owed to several individuals for insightful
  and helpful discussion as the article was being prepared, in
  particular: T. Lauer, C. Kankelborg, K. Kobayashi,
  V. Martinez-Pillet, and A. Parker. The author is also grateful to
  L. Golub for kindly providing NIXT data, and to the SDO imaging
  teams for the use of AIA and HMI data. Image analysis and processing
  were performed using the freeware Perl Data Language
  (http://pdl.perl.org).}

\bibliographystyle{plainnat}
\bibliography{gating}

\begin{thebibliography}{19}
\providecommand{\natexlab}[1]{#1}
\providecommand{\url}[1]{\texttt{#1}}
\expandafter\ifx\csname urlstyle\endcsname\relax
  \providecommand{\doi}[1]{doi: #1}\else
  \providecommand{\doi}{doi: \begingroup \urlstyle{rm}\Url}\fi

\bibitem[Bracewell(2000)]{bracewell_fourier_2000}
R.~N. Bracewell.
\newblock \emph{The {Fourier} {Transform} and {Its} {Applications}}.
\newblock 3rd edition, 2000.
\newblock ISBN 978-0-07-303938-1.

\bibitem[Buades et~al.(2005)Buades, Coll, and Morel]{buades_review_2005}
A.~Buades, B.~Coll, and J.~Morel.
\newblock A {Review} of {Image} {Denoising} {Algorithms}, with a {New} {One}.
\newblock \emph{Multiscale Model. Simul.}, 4\penalty0 (2):\penalty0 490--530,
  January 2005.
\newblock ISSN 1540-3459.

\bibitem[Davis(1989)]{davis_sound_1989}
G.~D. Davis.
\newblock \emph{The {Sound} {Reinforcement} {Handbook}}.
\newblock Yamaha, 1989.
\newblock ISBN 0-88188-900-8.

\bibitem[DeForest et~al.(2016)DeForest, Matthaeus, Viall, and
  Cranmer]{deforest_fading_2016}
C.~E. DeForest, W.~H. Matthaeus, N.~M. Viall, and S.~R. Cranmer.
\newblock \emph{Astrophys J}, 828\penalty0 (2):\penalty0 66, September 2016.
\newblock ISSN 1538-4357.

\bibitem[Golub et~al.(1990)Golub, Herant, Kalata, Lovas, Nystrom, Pardo,
  Spiller, and Wilczynski]{golub_sub-arcsecond_1990}
L.~Golub, M.~Herant, K.~Kalata, I.~Lovas, G.~Nystrom, F.~Pardo, E.~Spiller, and
  J.~Wilczynski.
\newblock Sub-arcsecond observations of the solar {X}-ray corona.
\newblock \emph{Nature}, 344\penalty0 (6269):\penalty0 842--844, April 1990.
\newblock ISSN 0028-0836.

\bibitem[Guo et~al.(2007)Guo, Au, Ma, and Liang]{guo_temporal_2007}
L.~Guo, O.~C. Au, M.~Ma, and Z.~Liang.
\newblock Temporal {Video} {Denoising} {Based} on {Multihypothesis} {Motion}
  {Compensation}.
\newblock \emph{IEEE Trans Circ \& Sys for Video Tech}, 17\penalty0
  (10):\penalty0 1423--1429, October 2007.
\newblock ISSN 1051-8215.

\bibitem[Huang(2015)]{huang_bayesian_2015}
Chao-Tsung Huang.
\newblock Bayesian {Inference} for {Neighborhood} {Filters} {With}
  {Application} in {Denoising}.
\newblock pages 1657--1665, 2015.

\bibitem[Jovanov et~al.(2009)Jovanov, Pizurica, Schulte, Schelkens, Munteanu,
  Kerre, and Philips]{jovanov_combined_2009}
L.~Jovanov, A.~Pizurica, S.~Schulte, P.~Schelkens, A.~Munteanu, E.~Kerre, and
  W.~Philips.
\newblock Combined {Wavelet}-{Domain} and {Motion}-{Compensated} {Video}
  {Denoising} {Based} on {Video} {Codec} {Motion} {Estimation} {Methods}.
\newblock \emph{IEEE Trans Circ \& Sys for Video Tech}, 19\penalty0
  (3):\penalty0 417--421, March 2009.
\newblock ISSN 1051-8215.

\bibitem[Kuan et~al.(1985)Kuan, Sawchuk, Strand, and
  Chavel]{kuan_adaptive_1985}
D.~T. Kuan, A.~A. Sawchuk, T.~C. Strand, and P.~Chavel.
\newblock Adaptive noise smoothing filter for images with signal-dependent
  noise.
\newblock \emph{IEEE Trans Pattern Anal Mach Intell}, 7\penalty0 (2):\penalty0
  165--177, February 1985.
\newblock ISSN 0162-8828.

\bibitem[Lebrun et~al.(2015)Lebrun, Colom, and Morel]{lebrun_noise_2015}
Marc Lebrun, Miguel Colom, and Jean-Michel Morel.
\newblock The {Noise} {Clinic}: a {Blind} {Image} {Denoising} {Algorithm}.
\newblock \emph{Image Processing On Line}, 5:\penalty0 1--54, January 2015.
\newblock ISSN 2105-1232.

\bibitem[Lee(1980)]{lee_digital_1980}
J.~S. Lee.
\newblock Digital {Image} {Enhancement} and {Noise} {Filtering} by {Use} of
  {Local} {Statistics}.
\newblock \emph{IEEE Trans Patt Anal \& Mach Intel}, PAMI-2\penalty0
  (2):\penalty0 165--168, March 1980.
\newblock ISSN 0162-8828.

\bibitem[Lemen et~al.(2011)Lemen, Title, Akin, Boerner, Chou, Drake, Duncan,
  Edwards, Friedlaender, Heyman, Hurlburt, Katz, Kushner, Levay, Lindgren,
  Mathur, McFeaters, Mitchell, Rehse, Schrijver, Springer, Stern, Tarbell,
  Wuelser, Wolfson, Yanari, Bookbinder, Cheimets, Caldwell, Deluca, Gates,
  Golub, Park, Podgorski, Bush, Scherrer, Gummin, Smith, Auker, Jerram, Pool,
  Soufli, Windt, Beardsley, Clapp, Lang, and Waltham]{lemen_atmospheric_2011}
James~R. Lemen, Alan~M. Title, David~J. Akin, Paul~F. Boerner, Catherine Chou,
  Jerry~F. Drake, Dexter~W. Duncan, Christopher~G. Edwards, Frank~M.
  Friedlaender, Gary~F. Heyman, Neal~E. Hurlburt, Noah~L. Katz, Gary~D.
  Kushner, Michael Levay, Russell~W. Lindgren, Dnyanesh~P. Mathur, Edward~L.
  McFeaters, Sarah Mitchell, Roger~A. Rehse, Carolus~J. Schrijver, Larry~A.
  Springer, Robert~A. Stern, Theodore~D. Tarbell, Jean-Pierre Wuelser, C.~Jacob
  Wolfson, Carl Yanari, Jay~A. Bookbinder, Peter~N. Cheimets, David Caldwell,
  Edward~E. Deluca, Richard Gates, Leon Golub, Sang Park, William~A. Podgorski,
  Rock~I. Bush, Philip~H. Scherrer, Mark~A. Gummin, Peter Smith, Gary Auker,
  Paul Jerram, Peter Pool, Regina Soufli, David~L. Windt, Sarah Beardsley,
  Matthew Clapp, James Lang, and Nicholas Waltham.
\newblock The {Atmospheric} {Imaging} {Assembly} ({AIA}) on the {Solar}
  {Dynamics} {Observatory} ({SDO}).
\newblock In Phillip Chamberlin, William~Dean Pesnell, and Barbara Thompson,
  editors, \emph{The {Solar} {Dynamics} {Observatory}}, pages 17--40. Springer
  US, 2011.
\newblock ISBN 978-1-4614-3672-0 978-1-4614-3673-7.

\bibitem[Liu and Freeman(2010)]{liu_high-quality_2010}
Ce~Liu and William~T. Freeman.
\newblock A {High}-{Quality} {Video} {Denoising} {Algorithm} {Based} on
  {Reliable} {Motion} {Estimation}.
\newblock In Kostas Daniilidis, Petros Maragos, and Nikos Paragios, editors,
  \emph{Computer {Vision} - {ECCV} 2010}, Lecture {Notes} in {Computer}
  {Science}, pages 706--719. Springer Berlin Heidelberg, September 2010.
\newblock ISBN 978-3-642-15557-4 978-3-642-15558-1.

\bibitem[Scherrer et~al.(2012)Scherrer, Schou, Bush, Kosovichev, Bogart,
  Hoeksema, Liu, Duvall, Zhao, Title, Schrijver, Tarbell, and
  Tomczyk]{scherrer_helioseismic_2012}
P.~H. Scherrer, J.~Schou, R.~I. Bush, A.~G. Kosovichev, R.~S. Bogart, J.~T.
  Hoeksema, Y.~Liu, T.~L. Duvall, J.~Zhao, A.~M. Title, C.~J. Schrijver, T.~D.
  Tarbell, and S.~Tomczyk.
\newblock The {Helioseismic} and {Magnetic} {Imager} ({HMI}) {Investigation}
  for the {Solar} {Dynamics} {Observatory} ({SDO}).
\newblock \emph{Sol Phys}, 275\penalty0 (1-2):\penalty0 207--227, January 2012.
\newblock ISSN 0038-0938, 1573-093X.

\bibitem[Schrijver et~al.(1997)Schrijver, Title, van Ballegooijen, Hagenaar,
  and Shine]{schrijver_sustaining_1997}
Carolus~J. Schrijver, Alan~M. Title, Adriaan~A. van Ballegooijen, Hermance~J.
  Hagenaar, and Richard~A. Shine.
\newblock Sustaining the {Quiet} {Photospheric} {Network}: {The} {Balance} of
  {Flux} {Emergence}, {Fragmentation}, {Merging}, and {Cancellation}.
\newblock \emph{Astrophys J}, 487:\penalty0 424--436, September 1997.
\newblock ISSN 0004-637X.

\bibitem[Shaick et~al.(2000)Shaick, Ridel, and Yaroslavsky]{shaick_hybrid_2000}
B.~Z. Shaick, L.~Ridel, and L.~Yaroslavsky.
\newblock A hybrid transform method for image denoising.
\newblock In \emph{2000 10th {European} {Signal} {Processing} {Conference}},
  pages 1--4, September 2000.

\bibitem[Shynk(1992)]{shynk_frequency-domain_1992}
J.~J. Shynk.
\newblock Frequency-domain and multirate adaptive filtering.
\newblock \emph{IEEE Signal Processing Magazine}, 9\penalty0 (1):\penalty0
  14--37, January 1992.
\newblock ISSN 1053-5888.

\bibitem[Walker et~al.(1988)Walker, Lindblom, Barbee, and
  Hoover]{walker_soft_1988}
Arthur B.~C. Walker, Jr., Joakim~F. Lindblom, Troy~W. Barbee, Jr., and
  Richard~B. Hoover.
\newblock Soft {X}-ray images of the solar corona with a normal-incidence
  {Cassegrain} multilayer telescope.
\newblock \emph{Science}, 241:\penalty0 1781--1787, September 1988.
\newblock ISSN 0036-8075.

\bibitem[Yaroslavsky(1996)]{yaroslavsky_local_1996}
Leonid~P. Yaroslavsky.
\newblock Local adaptive image restoration and enhancement with the use of
  {DFT} and {DCT} in a running window.
\newblock volume 2825, pages 2--13, October 1996.

\end{thebibliography}

\end{document}